\documentclass[12pt, prd, showpacs]{revtex4}
\usepackage{amssymb}

\input{tcilatex}

\begin{document}

\title{T-spheres as a limit of Lema\^{\i}tre-Tolman-Bondi solutions }
\author{O. B. Zaslavskii}
\affiliation{Department of Mechanics and Mathematics, Kharkov V.N.Karazin National
University, \\
Svoboda Square 4, Kharkov 61077, Ukraine}
\email{ozaslav@kharkov.ua}

\begin{abstract}
In the Tolman model there exist \ two quite different branches of solutions
- generic Lema\^{\i}tre-Tolman-Bondi (LTB) ones and T-spheres as a special
case. We show that, nonetheless, T-spheres can be obtained as a limit of the
class of LTB solutions having no origin and extending to infinity with the
areal radius approaching constant. It is shown that all singularities of
T-models are inherited from those of corresponding LBT solutions. In doing
so, the disc type singularity of a T-sphere is the analog of shell-crossing.
\end{abstract}

\maketitle

% It is always \today, today, but any date may be explicitly specified

%\keywords{Suggested keywords}
%Use showkeys class option if keyword display desired

The famous Tolman model \cite{tol} remains viable until now (see, for
example, \cite{void}, \cite{kras} for its astrophysical applications). Its
solutions split to two branches. The first one is the widely known Lema\^{\i}%
tre-Tolman-Bondi (LTB) solutions which describe an inhomogeneous collapse of
dust (or its time reversal). There is also one more brunch that arises as a
special solution of Einstein equations with the areal radius $R=R(t)$ not
depending on a spatial coordinate \cite{datt}, \cite{rub1}, \cite{rub2}
(recently such a type of solutions was discussed for rotating dust \cite%
{gair}). For such a kind of solution the 01 Einstein equation reduces to the
identity in contrast to the LTB case where its integration is essential for
finding the metric. The special solution under discussion (called
"T-spheres" or "T-models" in \cite{rub1}, \cite{rub2}) possesses a number of
unusual properties. For instance, they realize an "ideal gravitational
machine" in that an infinite amount of matter is bound to a finite mass,
etc. It was stressed in \cite{rub1}, \cite{rub2} that T-spheres cannot be
obtained from the LTB solutions. As far as the structure of singularities is
concerned, it was shown that some singularities of T-models are similar to
those of the closed Friedmann solution (which is the important particular
case of LTB solutions) but, in addition, there are also singularities of the
disk type.

The aim of the present work is to show that, in spite of the fact that
T-spheres are not contained in any LTB family of solutions, they can be
obtained from LTB as their limiting case. In doing so, the LTB prototype
that generates T-spheres is not the Friedmann solution but the solution that
contains no origin from one side. We also discuss the relationship between
singuilarities of both models and show that singularities of T-spheres
(including those that are absent in Friedmann-like models) are actually the
same as in the LTB prototype under discussion. Some time ago it was already
found that, by a quite different limiting transition, one can obtain from
the LTB family also the Vaidya one \cite{lem}, \cite{hel}, \cite{lem2} and
that the singularities of both models are the same. In this respect, our
results extend the similar relationship to T-models.

In general, the LTB solution is characterized by three function of which
only two (because of the freedom in rescaling a radial coordinate) can be
chosen arbitrarily$.$The metric can be written in the form

\begin{equation}
ds^{2}=-dt^{2}+\exp (2\omega )dr^{2}+R^{2}(t,r)d\Omega ^{2}\text{,}
\end{equation}%
\begin{equation}
\exp (2\omega )=\frac{R^{\prime 2}}{1+f(r)}\text{,}  \label{11}
\end{equation}%
where $d\Omega ^{2}=d\theta ^{2}+\sin ^{2}\theta d\phi ^{2}$, prime and dot
denote derivatives with respect to $r$ and $t$, correspondingly. For
simplicity, the cosmological term $\Lambda =0$ (but all results are
generalized easily to the case $\Lambda \neq 0$). Here%
\begin{equation}
\dot{R}^{2}=\frac{F(r)}{R}+f\text{,}
\end{equation}%
where $F(r)=2m(r)$, the quantity $m(r)$ plays the role of an active
gravitational mass. The function $f=f(r)$ can have any sign. For our
purposes, as will be seen from what follows, it is sufficient to restrict
ourselves to the case $f<0$ (the so-called elliptic case). Then

\begin{equation}
R=\frac{F}{2(-f)}(1-\cos \eta )\text{,}  \label{R}
\end{equation}%
\begin{equation}
\eta -\sin \eta =\frac{2(-f)^{3/2}}{F}(t-a)\text{, }a=a(r)\text{,}  \label{t}
\end{equation}%
$a(r)$ is the time of the big bang (if $t\geq a$) or\ big crunch (if $t\leq
a $). The energy density%
\begin{equation}
8\pi \rho =\frac{F^{\prime }}{R^{\prime }R^{2}}\text{.}
\end{equation}

We will need for what follows a convenient representation of $R^{\prime }$ 
\cite{dis} that can be found from (\ref{R}), (\ref{t}):%
\begin{equation}
R^{\prime }=(\frac{F^{\prime }}{F}-\frac{f^{\prime }}{f})R-[a^{\prime }+(%
\frac{F^{\prime }}{F}-\frac{3f^{\prime }}{2f})(t-a)]\dot{R}\text{.}
\label{R'}
\end{equation}

The solution for the T-model reads%
\begin{equation}
ds^{2}=-dt^{2}+b^{2}(t)dz^{2}+R(t)^{2}d\Omega ^{2}\text{,}  \label{t0}
\end{equation}%
\begin{equation}
R=\frac{R_{0}}{2}(1-\cos \eta )\text{, }\eta -\sin \eta =2\frac{(t-t_{0})}{%
R_{0}}\text{, }t_{0}=const\text{.}  \label{t1}
\end{equation}%
\begin{equation}
b(t)=\varepsilon \cot \frac{\eta }{2}+2M^{\prime }(r_{0})(1-\frac{\eta }{2}%
\cot \frac{\eta }{2})\text{,}  \label{b}
\end{equation}%
\begin{equation}
\rho =\frac{M^{\prime }(r_{0})}{4\pi R^{2}b(t)}\text{,}
\end{equation}%
where $M(r_{0})$ is the proper mass, $\varepsilon =0,\pm 1$.

Let the model have no origin on the left side, so that $R$ neither vanishes
nor grows unbounded but, instead, tends to a constant value at some $r=r_{0}$%
. To obtain the geodesically complete spacetime, we also assume that the
geometry approaches that of a semi-infinite throat, the function $f$ attains
its possible minimum value $f=-1$, so that near $r_{0}$

\begin{equation}
f=-1+k(r-r_{0})^{2}\text{, }k>0\text{. }  \label{fk}
\end{equation}

We also take $F^{\prime }$ to be finite at $r_{0}$. It is convenient to
rescale $r$ in such a way that $\left\vert a^{\prime }(r_{0})\right\vert =1$%
, if $a^{\prime }(r_{0})\neq 0$. Then it follows from (\ref{R}) - (\ref{R'})
that 
\begin{equation}
R(r_{0},t)=\frac{F(r_{0})}{2}(1-\cos \eta )\text{, }\dot{R}(r_{0})=\cot 
\frac{\eta }{2}\text{, }\eta -\sin \eta =\frac{2(t-a(r_{0}))}{F(r_{0})}
\end{equation}%
\begin{equation}
R^{\prime }(r_{0},t)=\varepsilon \cot \frac{\eta }{2}+F^{\prime }(r_{0})(1-%
\frac{\eta }{2}\cot \frac{\eta }{2})\text{.}
\end{equation}%
Here $\varepsilon =0,-1,1$ if $a^{\prime }(r_{0})=0,1,-1$ correspondingly.

(The particular explicit example of such a kind can be found in Sec. 6.4 of
Ref. \cite{dis} : $f=-1+B^{2}\exp (\frac{2r}{r_{0}})$, $F=A^{3}(1+C\exp [%
\frac{r}{r_{0}}])^{3}$, $a=0$. On the first glance, is looks different but
we can make the substitution $\exp (\frac{r}{r_{0}})=\tilde{r}-\tilde{r}_{0}$%
, where $\tilde{r}$ is the analog of our $r$ in previous formulas, whence it
is clear that it belongs to our family of solutions with $\varepsilon =0$
and $f$ having the form (\ref{fk}) everywhere, not only in the vicinity of $%
\tilde{r}_{0}$.)

The metric coefficient $g_{11}$ is equal to%
\begin{equation}
\exp [2\omega (r_{0})]=\frac{R^{\prime 2}(r_{0},t)}{k(r-r_{0})^{2}}\text{.}
\end{equation}

Let us make the substitution $r-r_{0}=A\exp (\alpha z)$, the constants are
chosen in such a way that $A>0$, $\alpha >0$, $\alpha ^{2}=k$, the new
variable $-\infty <z<\infty $. Then%
\begin{equation}
ds^{2}=-dt^{2}+b^{2}(t)dz^{2}+R^{2}(t)d\Omega ^{2}\text{,}
\end{equation}%
with 
\begin{equation}
b(t)=R^{\prime }(r_{0},t)\text{.}  \label{br'}
\end{equation}

But this is nothing other than the metric of the T-model. Although we
considered only the small vicinity of $r_{0}$, by substitution%
\begin{equation}
r=r_{0}+\lambda \chi \text{,}
\end{equation}%
where $\lambda \rightarrow 0$, $\chi $ is the new radial variable, after
repeating all steps we arrive at (\ref{t0}) - (\ref{b}) in \textit{all}
space. We should identify $t_{0}=a(r_{0})$. It remains to be seen that $%
F^{\prime }(r_{0})=2M^{\prime }(r_{0})$. Remembering that $F=2m=8\pi \int
drR^{2}R^{\prime }\rho $ and the proper mass $M=4\pi \int drR^{2}\exp (\frac{%
\omega }{2})\rho $, we have

\begin{equation}
M^{\prime }=\frac{F^{\prime }\exp (\omega )}{2R^{\prime }}\text{.}
\end{equation}

Taking into account that in our case lim$_{r\rightarrow r_{0}}\exp (\omega
)=b=R^{\prime }(r_{0},t)$, we see that indeed $F^{\prime }(r_{0})=2M^{\prime
}(r_{0})$ that completes our proof. Thus, T-models are indeed obtained as a
limit of LTB solutions.

Now we compare the singularities in both models. The non-vanishing
components of the Riemann tensor in the orthonormal frame are%
\begin{equation}
R_{\hat{0}\hat{\theta}\hat{0}\hat{\theta}}=\frac{\ddot{R}}{R}\text{,}
\end{equation}%
\begin{equation}
R_{\hat{r}\hat{0}\hat{r}\hat{0}}=\ddot{\omega}+\dot{\omega}^{2}\text{, }
\end{equation}%
\begin{equation}
R_{\hat{\theta}\hat{\phi}\hat{\theta}\hat{\phi}}=\frac{1}{R^{2}}[1-R^{\prime
2}\exp (-2\omega )+\dot{R}^{2}]\text{,}
\end{equation}%
\begin{equation}
R_{\hat{r}\hat{\theta}\hat{r}\hat{\theta}}=\frac{\dot{\omega}\dot{R}}{R}+%
\frac{\exp (-2\omega )}{R}(R^{\prime }\omega ^{\prime }-R^{\prime \prime })%
\text{.}
\end{equation}

To simplify comparison, let us use the variable $z$ instead of $r$ both in
T-models and LBT-solutions, $r-r_{0}=A\exp (\alpha z)$. Then (subscript
"LTB" and "T" refers to LTB and T-models, respectively) $\omega
_{LTB}\rightarrow \omega _{T}$, $R_{LTB}\rightarrow R_{T}$, $\left( \frac{%
\partial R}{\partial z}\right) _{LTB}\rightarrow 0=\left( \frac{\partial R}{%
\partial z}\right) _{T}$, $\left( \frac{\partial ^{2}R}{\partial z^{2}}%
\right) _{LTB}=\left( \frac{\partial R}{\partial r}\right) _{r_{0}}A\alpha
^{2}\exp (\alpha z)+\left( \frac{\partial ^{2}R}{\partial z^{2}}\right)
_{r_{0}}A^{2}\alpha ^{2}\exp (\alpha z)\rightarrow 0=\left( \frac{\partial
^{2}R}{\partial z^{2}}\right) _{T}$. Thus, all curvature components go
smoothly to those of T-models. As a result, the values of Kretsmann scalars
coincide and the singularities in both cases are the same.

It was observed by Ruban \cite{rub1}, \cite{rub2} that in T-models there are
singularities of the disc type ($b\rightarrow 0$, $R$ is finite). Now it is
clear that this type of singularities in the T-sphere solutions is inherited
from shell-crossing in LTB ones although the standard definition of
shell-crossing cannot be applied directly to T-spheres. Indeed, usually
shell-crossing occurs at some isolated values of $r$ where $R^{\prime }=0$, $%
F^{\prime }\neq 0$. In the T-case $R^{\prime }\equiv 0$, the coordinate $r$
becomes degenerate and all "shells" share the same value of $R$. If,
instead, one labels shells in the T-case by values of $z$, the condition $%
b=0 $ means just crossing of shells with different $z$.

Thus, the Vaidya spacetime and T-spheres can be obtained from the Tolman
models by two different transitions. In the first case $f\rightarrow +\infty 
$ \cite{lem}, \cite{hel}, \cite{lem2}, in the second one $f\rightarrow -1$
according to (\ref{fk}). In both cases the singularities of the LTB
prototype and its limiting counterpart coincide. From the general viewpoint,
such limiting transitions represent examples of the "limits of spacetime" 
\cite{ger}. It would be interesting to elucidate, whether such a
relationship between singularities is inherent to this procedure as such or
it is rather the property of some particular models.


\begin{thebibliography}{99}
\bibitem{tol} R. C. Tolman, 1934, Proc. Nat. Acad. Sci. USA, 20, 169;
reprinted in 1997, Gen. Relativ. Gravitation, 29, 935

\bibitem{void} K. Bolejko, A. Krasinski and C.Hellaby, Formation of voids in
the Universe within the Lemaitre-Tolman model, gr-qc/0411126.

\bibitem{kras} A. Krasinski and C.Hellaby, Phys.Rev. D69 (2004) 043502.

\bibitem{datt} Datt, B. (1938). Z. Physik 108, 314 [Reprinted: Gen. Rel.
Grav. 31, 1619 (1999)].

\bibitem{rub1} V. A. Ruban, Sov. Phys. JETP (Letters) 8\textbf{,} 414
(1968). [Reprinted: Gen. Rel. Grav. 33\textbf{,} 369 (2001)].

\bibitem{rub2} V. A. Ruban, Sov. Phys. JETP 29\textbf{,} 1027 (1969).
[Reprinted: Gen. Rel. Grav. 33, 375 (2001)].

\bibitem{gair} J. R. Gair, Class. Quant. Grav. 19, 6345 (2002).

\bibitem{lem} J. P. S. Lemos, Phys. Rev.\ Lett. 68 (1992) 1447.

\bibitem{hel} C. Hellaby, Phys.\ Rev. D 49 (1994) 6484.

\bibitem{lem2} S. Gao and J. P. S. Lemos, Phys. Rev. D 71 (2005) 084022.

\bibitem{dis} C. Hellaby, Some properties of singularities in the Tolman
model, Ph. D. Theses, 1985. Queen's University at Kingston, Ontario, Canada.

\bibitem{ger} Geroch R Commun. Math. Phys. 13 (1969 ) 180.
\end{thebibliography}
\end{document}